\documentclass{aa}

\usepackage{graphicx}
\usepackage{txfonts}
\usepackage{natbib}	
\usepackage{xcolor} 

\bibpunct{(}{)}{;}{a}{}{,} 

\begin{document}

   \title{High resolution transmission spectrum of the Earth's atmosphere}

   \subtitle{Seeing Earth as an exoplanet using a lunar eclipse}

   \author{F. Yan
          \inst{1,2,3}
          \and
          R. A. E. Fosbury
          \inst{3}
          \and
          M. G. Petr-Gotzens
          \inst{3}
          \and
          G. Zhao
          \inst{1}
          \and
          W. Wang
          \inst{1}
          \and
          L. Wang
          \inst{1}
          \and
          Y. Liu
          \inst{1}
          \and
          E. Pall\'e
          \inst{4,5}
		}

   \institute{Key Laboratory of Optical Astronomy, National Astronomical Observatories, Chinese Academy of Sciences, 20A Datun Road, Chaoyang District, 100012 Beijing, China\\
\email{feiy@nao.cas.cn, gzhao@nao.cas.cn}
         \and
         	University of Chinese Academy of Sciences, 19A Yuquan Road, Shijingshan District, 100049 Beijing, China
         \and
			European Southern Observatory, Karl-Schwarzschild-Str. 2, 85748, Garching bei M\"unchen, Germany
		\and
			Instituto de Astrof\'isica de Canarias, C/ v\'ia L\'actea, s/n, 38205 La Laguna, Tenerife, Spain
        \and			
			Dpto. de Astrof\'isica, Universidad de La Laguna, 38206 La Laguna, Tenerife, Spain
\\	
         }
	\date{Received -----; accepted -----}


  \abstract
  {With the rapid developments in the exoplanet field, more and more terrestrial exoplanets are being detected. Characterising their atmospheres using transit observations will become a key datum in the quest for detecting an Earth-like exoplanet.
The atmospheric transmission spectrum of our Earth will be an ideal template for comparison with future exo-Earth candidates. By observing a lunar eclipse, which offers a similar configuration to that of an exoplanet transit, we have obtained a high resolution and high signal-to-noise ratio transmission spectrum of the Earth’s atmosphere.
This observation was performed with the High Resolution Spectrograph at Xinglong Station, China during the total lunar eclipse in December 2011.
We compare the observed transmission spectrum with our atmospheric model, and determine the characteristics of the various atmospheric species in detail.
In the transmission spectrum, $\mathrm{O_2}$, $\mathrm{O_3}$, $\mathrm{O_2 \cdot O_2}$, $\mathrm{NO_2}$ and $\mathrm{H_2O}$ are detected, and their column densities are measured and compared with the satellites data. 
The visible Chappuis band of ozone produces the most prominent absorption feature, which suggests that ozone is a promising molecule for the future exo-Earth characterization.
Due to the high resolution and high signal-to-noise ratio of our spectrum, several novel details of the Earth atmosphere's transmission spectrum are presented.
The individual $\mathrm{O_2}$ lines are resolved and $\mathrm{O_2}$ isotopes are clearly detected. 
Our new observations do not confirm the absorption features of Ca\,II or Na\,I which have been reported in previous lunar eclipse observations. However, features in these and some other strong Fraunhofer line positions do occur in the observed spectrum. We propose that these are due to a Raman-scattered component in the forward-scattered sunlight appearing in the lunar umbral spectrum.
Water vapour absorption is found to be rather weak in our spectrum because the atmosphere we probed is relatively dry, which prompts us to discuss the detectability of water vapour in Earth-like exoplanet atmospheres.
}

   \maketitle
%

\section{Introduction}

Almost 1800 exoplanets have been detected so far\footnote{see http://exoplanet.eu/catalog.php, updated on May 9, 2014.}. Among them, 1133 are transiting planets, which allow follow-up studies of their atmospheres during primary transit and secondary eclipse. Restricted by detection limits of current available instruments, most of such studies were done for hot Jupiters \citep[e. g. ][]{Charbonneau2002,Wang2013}. For a review on the atmospheres of exoplanets, please read \cite{Seager2010a}.
This subject of the characterization of exoplanets developed quickly after this review, including a dramatic increase of the number of planets with their atmospheres been studied, and massive studies on the super-Neptune GJ 436b, and the super-Earth GJ 1214b in both observational and theoretical aspects 
\citep[e. g. ][]{Charbonneau2009,Howe2012,Kreidberg2014}.


After the first announcement of the discovery of two Earth-sized exoplanets (Kepler-20e and Kepler-20f) by \cite{Fressin2012}, nearly a hundred Earth-sized or smaller exoplanets have now been detected. 
\cite{Dumusque2012} discovered an Earth-mass planet  orbiting $\alpha$ Centauri B -- a member of the Sun's closest neighbouring system. 
Recently, \cite{Quintana2014} detected an Earth-sized exoplanet in the Habitable Zone of its host star.
The statistics of detected planetary systems and the knowledge of the selection effects indicate that the planet occurrence rate increases toward smaller planet radii \citep{Howard2012}, which means rocky planets are expected to be common in the universe.
It is of great astrophysical interests and importance to study their atmospheres and habitabilities, however, it might only be done in the future using the next generation telescopes such as E-ELT \citep{Hedelt2013}.

Appropriate observations of the Earth's atmosphere can be used to guide the exoplanet atmosphere models and refine observing strategies, providing valuable insights into the future exo-Earth atmosphere characterisation. Lunar eclipses provide unique opportunities to acquire the transmission spectrum of the Earth's atmosphere in an appropriate geometry. Therefore, we observed the total lunar eclipse in December 2011 with a high resolution spectrograph to obtain the high resolution and high 
signal-to-noise ratio (SNR) transmission spectrum of the Earth's atmosphere.

Several previous investigations have paved the way for this work.
\citet[hereafter P09]{Palle2009} observed a partial lunar eclipse in August 2008 and acquired the transmission spectrum from the umbral spectra with a resolving power of R $\sim$ 960. They detected $\mathrm{O_2}$, $\mathrm{O_3}$, $\mathrm{H_2O}$, $\mathrm{O_2 \cdot O_2}$, $\mathrm{Ca\, II}$, $\mathrm{Na\, I}$ in the optical as well as $\mathrm{CH_4}$, $\mathrm{CO_2}$ and $\mathrm{O_2 \cdot N_2}$ in the infrared.
\cite{Vidal2010} observed the same partial lunar eclipse and found $\mathrm{O_2}$, $\mathrm{O_3}$ and $\mathrm{Na\, I}$ based on the transmission spectrum obtained from the penumbral spectra.
\citet[hereafter U13]{Ugolnikov2013} observed the total lunar eclipse in December 2011  with a resolving power of R $\sim$ 30,000 and a SNR of 45 near the $\mathrm{H_\alpha}$ line and quantitatively analysed the atmospheric species, including $\mathrm{O_2}$, $\mathrm{O_3}$, $\mathrm{H_2O}$, $\mathrm{O_2 \cdot O_2}$ and $\mathrm{NO_2}$.

The transmission spectrum obtained from our umbral eclipse observation is of high resolution and high SNR compared to the previous results, which is a significant improvement in the study of Earth's transmission spectrum. Several novel details of the Earth's transmission spectrum are presented in this paper. These include the clear detection of oxygen isotopes; the confirmation of the non-detection of Na\,I or Ca\,II absorptions which have been reported in previous lunar eclipse observations; the `emission' features at the Fraunhofer lines which are regarded as the indication of Raman scattering in the forward-scattered skylight component. 
The water vapour absorption is found to be weak in our spectrum compared to that in P09's spectrum, and we further discuss the water vapour detectability in exo-Earth atmospheres.
The quantitative calculation of $\mathrm{O_2 \cdot O_2}$ absorption bands also shows differences with previous results, and so several different factors which affect the $\mathrm{O_2 \cdot O_2}$ absorption bands measurement are discussed.


The paper is organised as follows. Sect. 2 is a description of our observations. In Sect. 3, the transmission spectrum is presented and analysed. In Sect. 4, a one-dimensional telluric atmospheric model is built with which the atmospheric species are quantitatively calculated. In Sect. 5 and 6, we present the discussion and a summary.



\section{Observations and data reduction}
\begin{figure}
   \centering
   \includegraphics[width=0.550\textwidth]{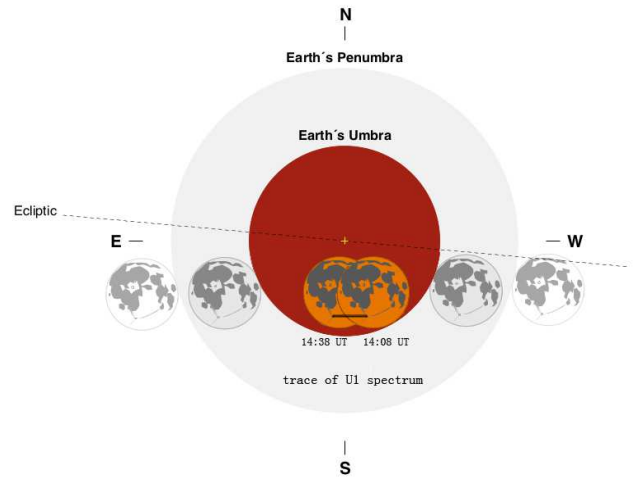}
      \caption{A sketch of the total lunar eclipse on 10 December 2011. The black solid line shows the trace of the fibre location during the observation of the umbra spectrum. The figure is reproduced from the NASA eclipse page (http://eclipse.gsfc.nasa.gov/eclipse.html).}
         \label{NASA}
   \end{figure}

The total lunar eclipse took place on 10 December 2011 (Figure \ref{NASA}). The observations were made with the fibre-fed echelle High Resolution Spectrograph (HRS) mounted on the 2.16-m telescope at Xinglong Station, China.
The slit width was set to 190 $\mu$m, corresponding to a resolving power ($\lambda / \Delta \lambda$) of R $\sim$ 45,000 with 3.2 pixel sampling. The single 4K $\times$ 4K CCD covers the wavelength region of 4300 $\sim$ 10000$\,\AA$.
The telescope was  pointed to the same part of the lunar surface --- near the Tycho crater where the albedo is high --- for all the exposures.

The two lunar spectra used for this study were taken when the Moon was in the umbra and out of the Earth's shadow (which is referred to as the bright Moon), respectively. The details of the umbra spectrum (U1) and the bright Moon spectrum (B2) are shown in Table \ref{observation}. The airmass of U1 changed from 1.10 to 1.16 during the 30 minutes observation while the airmass of B2 was 1.17 during the 6 seconds observation.

The standard data reductions, including bias correction, order definition, flat fielding and background subtraction, were performed using IRAF \footnote{IRAF is distributed by the National Optical Astronomical Observatory, which is operated by the Association of Universities for Research in Astronomy, Inc., under cooperative agreement with the National Science Foundation}
and self-written Python scripts.
The wavelength calibration was performed by comparing the positions of the emission lines of the Th-Ar halogen lamp and has a typical uncertainty of 0.3 km/s.
Flux calibration was not necessary, since we only used the ratio of the umbral to the bright Moon for the analysis. The one-dimensional spectra were extracted and all of the 77 spectral orders were combined.
The SNR for U1 is $\sim$ 500 at 8000$\,\AA$ with a pixel size of 0.035$\,\AA$ and is $\sim$ 60 at 5000$\,\AA$ with a pixel size of 0.022$\,\AA$.


%
\begin{table*}
\caption{Detail information of the observed spectra.}             
\label{observation}      
\centering                          
\begin{tabular}{c c c c c}        
\hline\hline                 
ID	&	Observation start time (UT)	&	Observed position 	&	Exposure time (s)	&	Average airmass \\    
\hline                        
 		U1	&	14:08:25	&	Umbra	&	1800	&	1.13	\\
        B2	&	18:10:24	&	Bright Moon	&	6	&	1.17	\\
\hline                                   
\end{tabular}
\end{table*}


\section{Data analysis}
\subsection{Transmission spectrum determination}

\begin{figure}
   \centering
   \includegraphics[width=0.5\textwidth]{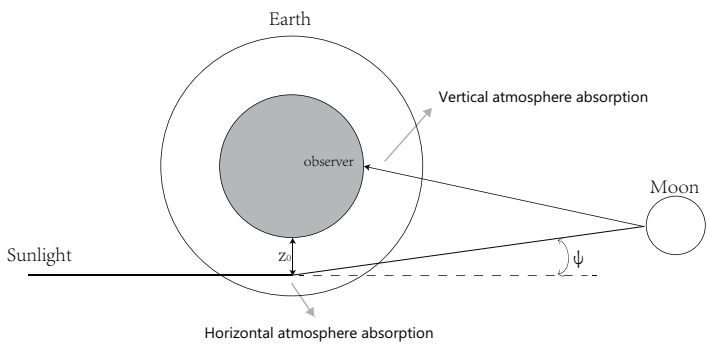}
      \caption{ A schematic diagram of the lunar eclipse. The atmospheric thickness and the refraction angle $\psi$ have been exaggerated. Some small contribution will arise from forward-scattering by the illuminated sky at the terminator.
    }
         \label{eclipsing}
   \end{figure}

During a total lunar eclipse, the Moon moves into the Earth's umbra and the light reflected from the lunar surface is predominantly the sunlight refracted by the Earth's atmosphere (as shown in Figure \ref{eclipsing}) with some small contribution arising from forward-scattering by the illuminated sky \citep{Garcia2011a}.
Figure \ref{eclipsing} indicates that the umbra spectrum contains two components of atmospheric absorption: 1) horizontal (long pathlength) absorption of the atmosphere through the day-night terminator which is similar to the situation in exoplanet transits; and 2) near-vertical absorption by the local atmosphere above the telescope. For the bright Moon spectrum, the atmospheric absorption only has the latter.

\begin{figure*}
   \centering
   \includegraphics[width=0.98\textwidth]{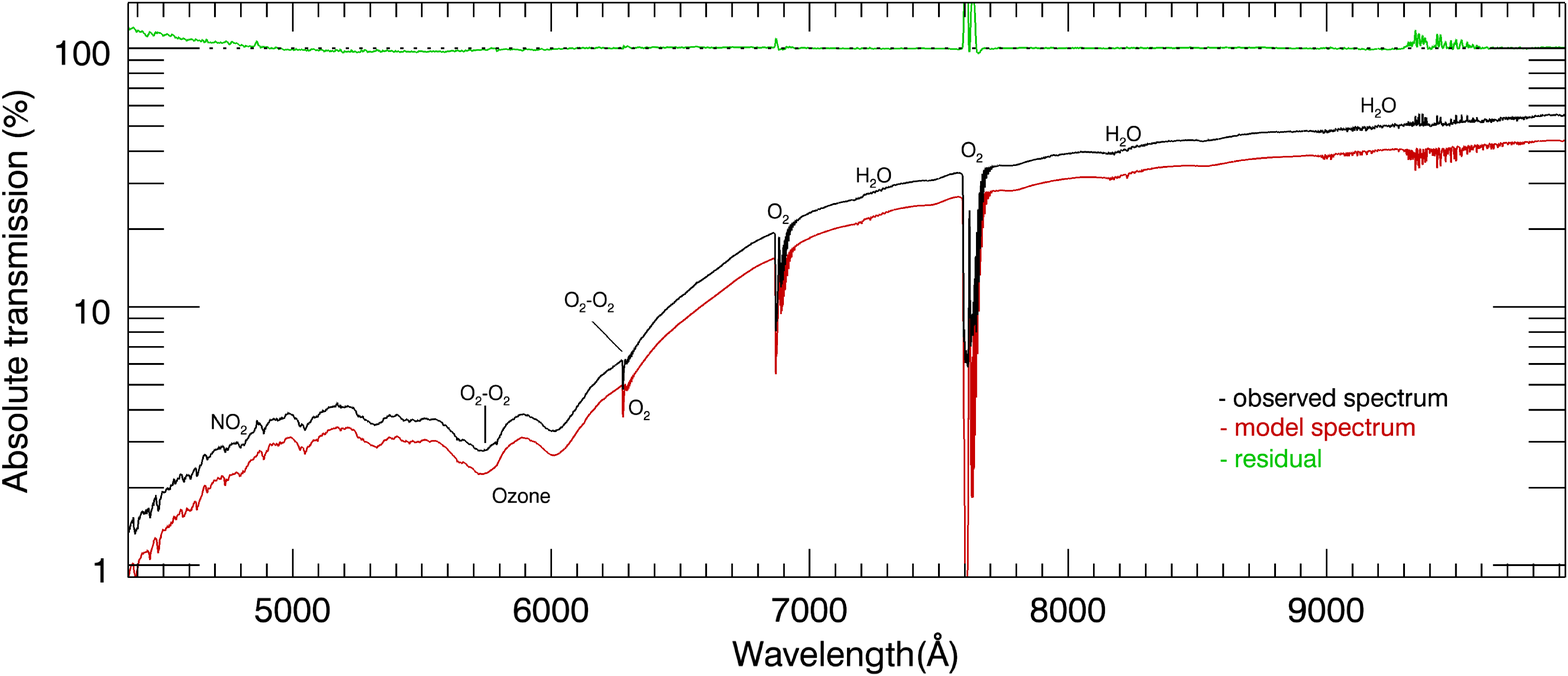}
\caption{The observed transmission spectrum of the Earth's atmosphere together with the model spectrum. The black solid line is the observed spectrum binned every 50 pixels. The red solid line is the model spectrum. The model spectrum is shifted down 20\% for clarification and the green line at the top shows the residual between observed and model spectra.
The oxygen lines around 7600 $\,\AA$ and the water lines around 9500 $\,\AA$ do not fit well because of absorption saturation in the observed spectrum.
}
         \label{final-model}
   \end{figure*}

In order to obtain the Earth's atmospheric transmission spectrum, we remove the vertical absorption by performing the ratio of U1 to B2 under the assumption that the amount of the vertical absorption keeps the same in the two spectra.
This is generally true, because the vertical airmass difference between U1 and B2 is merely 0.04, while the tangent airmass is about $10\sim40$ in U1. Therefore the slight airmass difference can be safely ignored for the transmission spectrum determination, and this will not affect the quantitatively calculation for most of the atmospheric species except for $\mathrm{H_2O}$ (cf. Section 4.2.2 for details).
The ratio also cancels out the solar spectrum features, the effect of the lunar surface spectral albedo and the intensity-wavelength response of the spectrograph. The transmission spectrum calculated with this method is displayed in Figure \ref{final-model}.

Figure \ref{Ha} shows the spectra around the Fraunhofer C-line (H$\alpha$). The figure demonstrates that the solar spectral features are well removed in the transmission spectrum. The slight structure at the C-line position is due to the different radial velocities (RVs) of the spectra (see Section 3.3).

%

%

  %

\subsection{Geometric calculation}
In order to interpret the transmission spectrum, we need to establish which part of the Earth's atmosphere contributes to the refracted sunlight at the observed umbra position. The Geometric Pinhole Model \citep{Vollmer2008} is applied to calculate the altitude in the atmosphere at the tangent point. This model assumes the terminator of the atmosphere is consisted of different tangent pinholes with minimum altitude $z_\mathrm{0}$ and refraction angle $\psi$ (as shown in Figure \ref{eclipsing}). As the observation proceeds for a specific position on the lunar surface, the position and minimum altitude of the pinhole forms a 2D locus at the Earth's terminator. This 2D locus is the refracted solar image as seen from the lunar surface.



\begin{figure}
   \centering
   \includegraphics[width=0.45\textwidth]{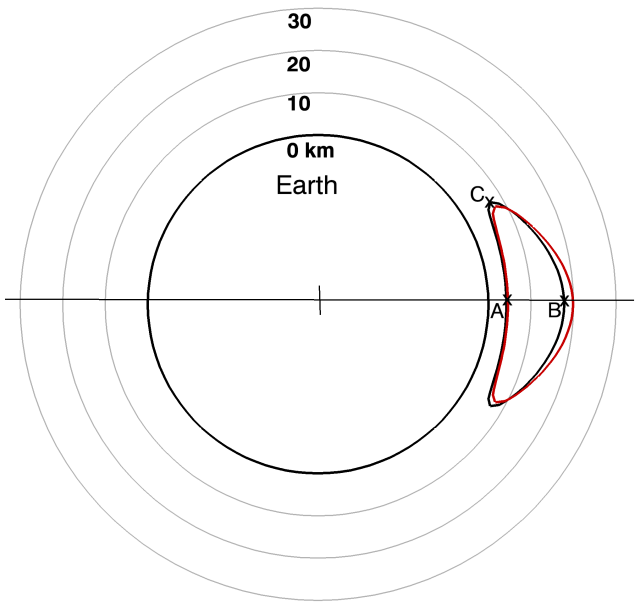}
\caption{The solar image as observed from the lunar surface during the umbra observation. This image is the altitude range for the pinholes at the Earth's terminator. The solid black line is the range when the Tycho crater has the smallest distance to the umbral center. A and B are the lower and upper point (4.2, 17.8 km), respectivley, and C represents the the altitude at the edge of the range (7.1 km). The red line is the range when the distance between Tycho and umbra center is the largest. The thickness of the atmosphere is exaggerated.}
         \label{altitude}
   \end{figure}

During the observation of U1, Tycho's position in the umbra is changing, which makes the 2D image change. For spectrum U1, the altitude range is 4.2 $\sim$ 17.8 km for the minimum distance between Tycho's position and the umbra center, and is 4.6 $\sim$ 20.0 km for the maximum distance. Figure \ref{altitude} shows the 2D images of the sun, in which the variation of the altitude range over time is not significant during the umbra exposure.

\begin{figure}
   \centering
   \includegraphics[width=0.45\textwidth]{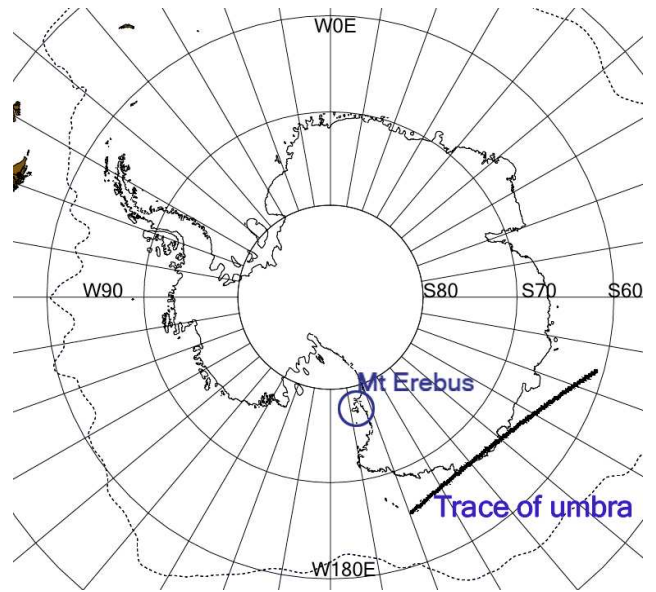}
\caption{The geometric trajectory of the tangent-point for the transmission spectrum. The geographical map is taken from Australian Antarctic Data Centre. The position of Mt Erebus is circled. }
         \label{coordinate}
   \end{figure}

The geometric longitude and latitude at the low altitude point (point A in Figure \ref{altitude}) are calculated with the JPL Ephemeris\footnote{Http://ssd.jpl.nasa.gov/?ephemerides}. 
Figure \ref{coordinate} shows the coordinate trajectory during U1 exposure, which is above the coast of Antarctic Ocean with the mid-point at $156^\circ$E, $67^\circ$S.


An overall picture of the transmission geometry is described in the following: the path of the sunlight through the Earth's terminator are above the Antarctica Ocean, with an altitude range of 4 $\sim$ 20 km.


\subsection{Radial velocity correction}

\begin{figure}
   \centering
   \includegraphics[width=0.5\textwidth]{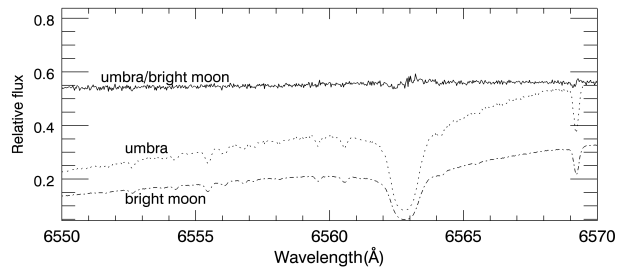}
\caption{The spectra around the H$\alpha$ line. The dotted and dot-dashed lines represent the umbra spectrum and the bright Moon spectrum, respectively. The solid line is the ratio of umbra to the bright Moon spectrum. }
         \label{Ha}
   \end{figure}

The detailed radial velocity in the transmission spectrum is complicated due to the motion of the Earth-Moon system.
For the solar Fraunhofer lines, the RVs are the combination of Earth’s rotation, orbital motion, and the lunar motion components, and the RV difference between U1 and B2 is 0.4 km/s.
For the vertical Earth atmospheric absorption lines, however, the RVs are all zero. In order to eliminate the vertical atmospheric absorption in the transmission spectrum as  much as possible, we did not correct the RVs of the Fraunhofer lines. 
This non-correction of RV will only slightly affect our transmission spectrum at the Fraunhofer line positions as shown in Figure \ref{Ha}. Therefore, our quantitative calculations in Section 4 will not be altered.

Besides, the entire transmission spectrum exhibits a non-zero radial velocity, which corresponds to the RV of the horizontal atmosphere.
According to the theoretical calculation with JPL Ephemeris, the RV is 0.35 km/s. Considering that the Earth's atmospheric absorption lines in B2 have a zero RV, we used B2 spectrum as a template to perform a cross-correlation between the transmission spectrum and B2 spectrum and got the RV of the oxygen lines as 0.45 km/s. This result is considered in the quantitative calculations in Section 4.

%

\subsection{Molecular features}

The atmospheric spectral features are marked in Figure \ref{final-model}. Two of the principal contributors to the continuous spectrum are due to the molecular Rayleigh scattering and aerosol scattering.

The most prominent sharp features in the transmission spectrum are the results of the $\mathrm{O_2}$  lines absorption in the three $\mathrm{O_2}$ bands (around 0.63 $\mathrm{\mu m}$, 0.69 $\mathrm{\mu m}$, 0.76 $\mathrm{\mu m}$). 
Thanks to the high resolution of our spectra, the $\mathrm{O_2}$ lines are clearly resolved and modelled in Sect. 4.

The $\mathrm{H_2O}$ lines are very weak in our transmission spectrum which results, presumably, from the well-known  low water vapour  concentration in the atmosphere above Antarctica. Water vapour detectability  will be discussed in Sect. 5.

\begin{figure*}
   \centering
   \includegraphics[width=0.98\textwidth]{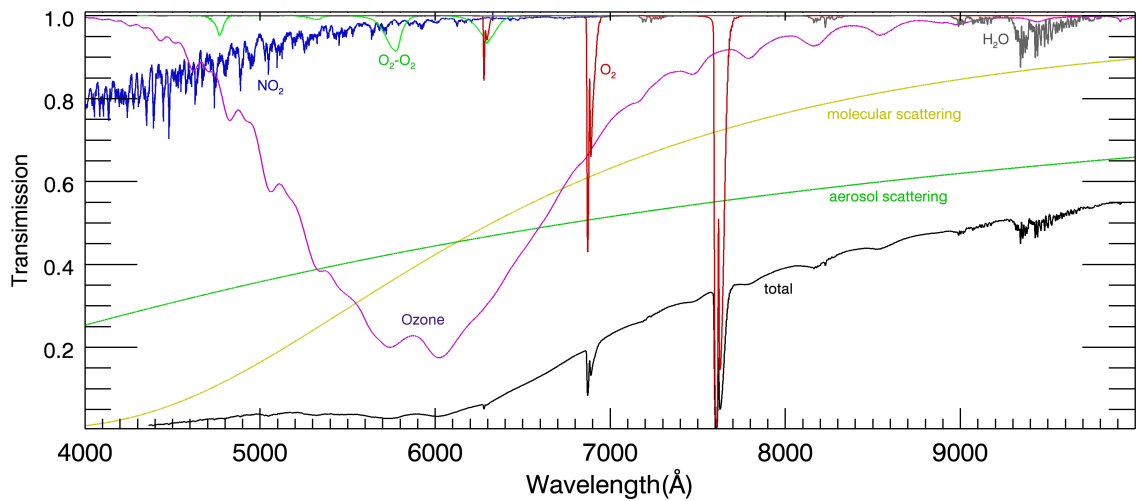}
\caption{Contributions of the different species used in the model spectrum.
The figure clearly shows that ozone is the main absorption source between 5500 $\,\AA$ and 6500 $\,\AA$, while $\mathrm{NO_2}$ dominates the spectral feature at the blue end (< 5000 $\,\AA$). }
         \label{species}
   \end{figure*}

Ozone absorption in the Chappuis band is strong and acts as the main absorption source between 0.55 $\mathrm{\mu m}$ and 0.65 $\mathrm{\mu m}$ (cf. Figure \ref{species}). The absorption is so significant that it leads to obvious modifications of the eclipsed Moon's color \citep{Fosbury2011}.

Spectral features from the trace gas $\mathrm{NO_2}$ are visible at the blue end of the spectrum. 
Since much of the blue light has been scattered, only observations with a high SNR like ours are capable to detect $\mathrm{NO_2}$. 

The $\mathrm{O_2 \cdot O_2}$ absorption in the Earth's atmosphere is due to the collision induced absorption (CIA) of two $\mathrm{O_2}$ molecules \citep{Thalman2013}. $\mathrm{O_2}$ CIA may become of significant interest for terrestrial exoplanet studies since its pressure dependence can give information about the lower atmosphere \citep{Misra2013arXiv}. In our spectrum, the three $\mathrm{O_2 \cdot O_2}$ CIA bands (around 477 nm, 577 nm and 630 nm) are clearly detected.


On the other hand, we do not detect any atomic or ionic absorption features (such as Ca\,I, Ca\,II and Na\,I). 
They are expected to be mainly presented  in the upper mesosphere \citep[see][]{Plane1999,Gerding2001}.
We do, however, detect apparent `emission' features in some strong Fraunhofer line positions  in our transmission spectrum, which we propose are due to Raman scattering in the forward-scattered sunlight (see section 5.3 for a detailed discussion).


%

\section{Model and quantitative calculation}

\subsection{Model description}
In the present work, a line-by-line one dimensional model is built, which contains $\mathrm{O_2}$, $\mathrm{H_2O}$, $\mathrm{O_3}$,  $\mathrm{NO_2}$, $\mathrm{O_2 \cdot O_2}$ , Rayleigh and aerosol scattering.
The least-square method is used to derive the best fit to the obtained transmission spectrum, as shown in
Figure \ref{final-model}. In Figure \ref{species}, contributions of various species included in the model are shown. 


\subsubsection{Absorption of atmospheric species}
The transmission spectrum is calculated based on the Beer–Lambert law:
\begin{equation}
      T\,=\,e^{-\sum (u\, \sigma)} ,
   \end{equation}
where $T$, $u$ and $\sigma$ are transmission, column density and cross section of each atmospheric species, respectively.

The cross sections of ozone and $\mathrm{NO_2}$ are taken from ``Molecular Spectroscopy'' at IUP Bremen\footnote{Http://www.iup.physik.uni-bremen.de/gruppen/molspec/index.html}. The $\mathrm{O_2 \cdot O_2}$ absorption data are taken from GEISA database \citep{Jacquinet2011}.
The $\mathrm{O_2}$ and $\mathrm{H_2O}$ line lists are taken from HITRAN database \citep{Rothman2009}, and the absorption cross sections are calculated line-by-line for each given temperature and pressure with Lorentz line profiles.

The atmospheric scattering consists of two parts: molecular scattering and aerosol scattering. The molecular Rayleigh scattering cross section for the Earth's atmosphere is taken from \cite{Bodhaine1999} :
   \begin{equation}
      \sigma_{\mathrm{Rayleigh}} = \frac{1.0456 - 341.29\lambda^{-2} - 0.9023\lambda^{2}}{1+0.002706\lambda^{-2} - 85.9686\lambda^{2}} \,\mathrm{(\times 10^{-28}cm^2)}\,,
   \end{equation}
where $\mathrm{\lambda}$ is the vacuum wavelength in $\mathrm{\mu m}$. According to \cite{Bucholtz1995} this Rayleigh scattering cross section is independent of temperature and pressure.

For the aerosol scattering, we assume that the aerosols are distributed uniformly in the atmosphere and assign the cross section per molecule \citep{Kaltenegger2009}. The aerosol scattering cross section data for the Earth's atmosphere is then taken from \cite{Allen1976} and expressed as:
\begin{equation}
      \sigma_{\mathrm{aerosol}} = 1.4\times 10^{-27} \lambda^{-1.3}\,\mathrm{(cm^2)}\,,
   \end{equation}
where $\sigma_{\mathrm{aerosol}}$ is the aerosol absorption cross section corresponding to each air molecule. As this equation is suitable for normally clear atmospheric conditions, we introduce a factor $A_{\mathrm{aerosol}}$ which allows adjustment with respect to `clear' conditions.

   \begin{figure}
   \centering
   \includegraphics[width=0.5\textwidth]{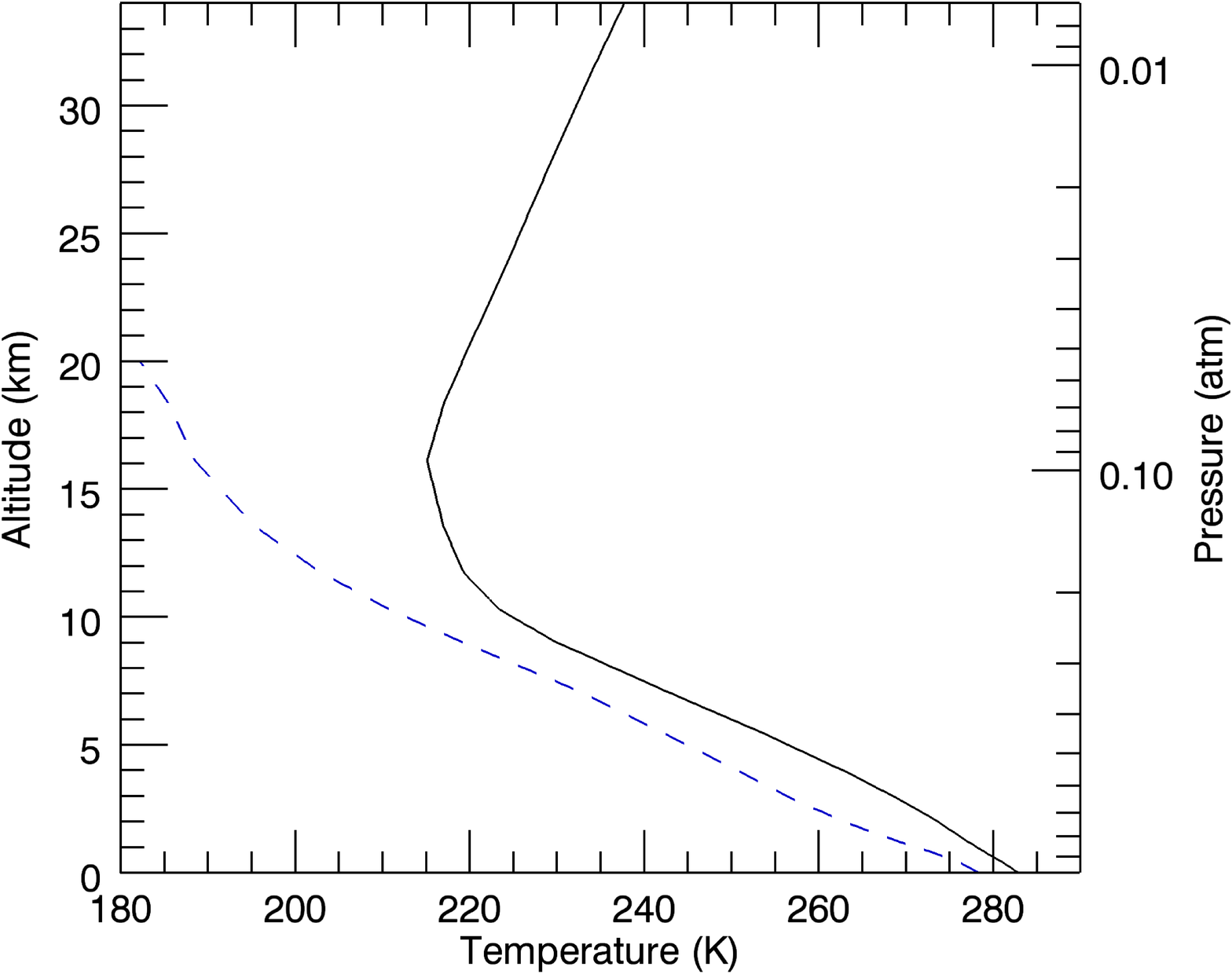}
   \includegraphics[width=0.45\textwidth]{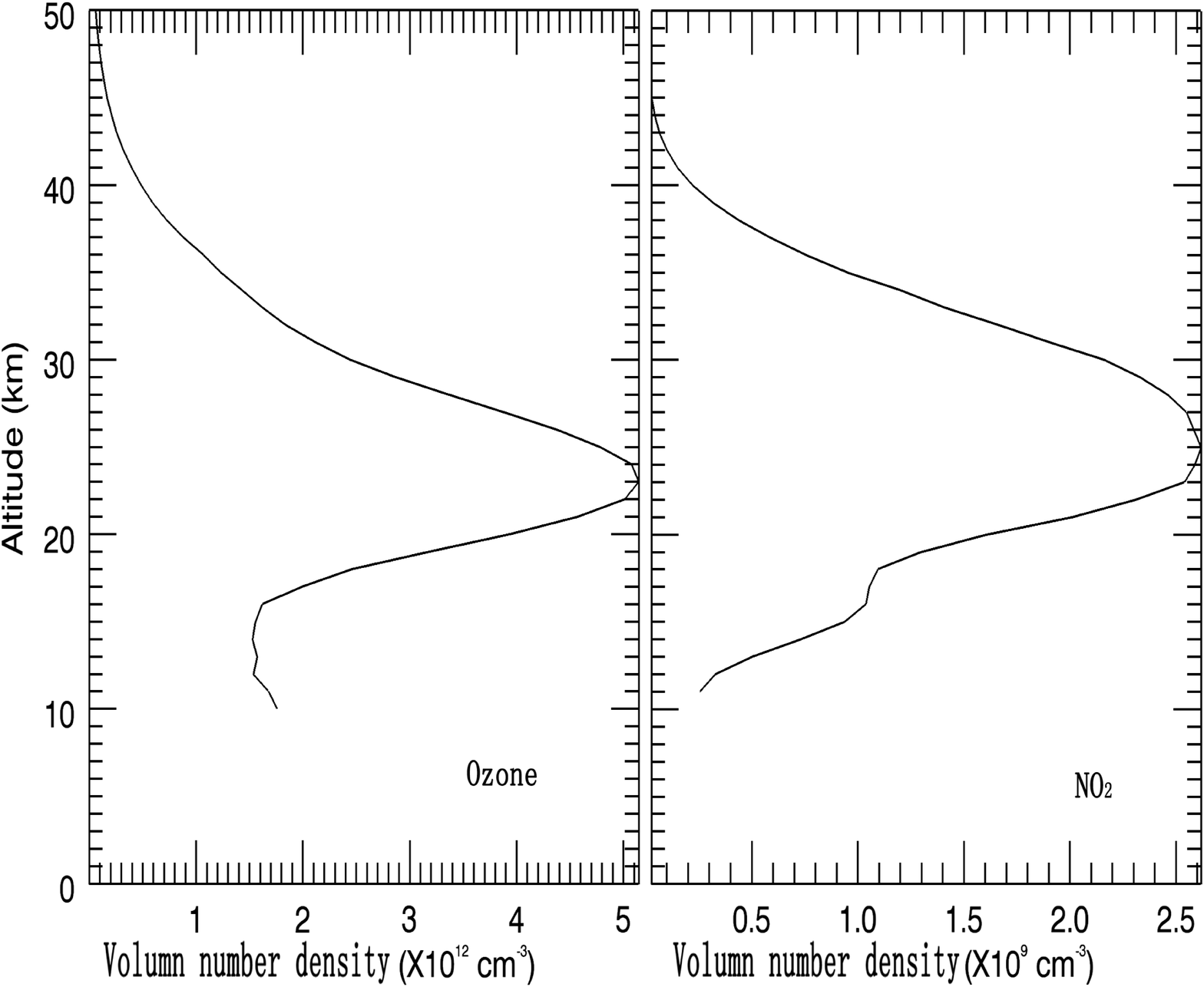}
\caption{The atmosphere profiles above the trajectory shown in Figure \ref{coordinate}. The top figure shows the T-P profile (black solid line) and the dew-point temperature (blue dashed line) which are taken from the MODIS MOD07\_ L2 data. The bottom figure shows the ozone and $\mathrm{NO_2}$ volume density profile which are taken from the IUP/IFE-UB SCIAMACHY data.
}
         \label{TPprofile}
   \end{figure}
%

\subsubsection{Model geometry and atmospheric profiles}
As shown in Figure \ref{altitude}, the light of the transmission spectrum comes from a specific atmospheric altitude range. In order to simplify our model, we use the `effective altitude' instead of the altitude range to represent the mean atmosphere property for the observed transmission spectrum.

For a light path with the minimum height $z_\mathrm{0}$ (as indicated in Figure \ref{eclipsing}), the optical depth for each molecular species is integrated along the path under the assumption that the absorption cross section $\sigma$ does not vary with the temperature and pressure:
\begin{equation}
\tau = u \cdot \sigma = \int n_z  \, \mathrm{d} z \cdot \sigma ,
   \end{equation}
where $n_z$ is the volume density of the molecular species at different altitude $z$.
The small effect of the atmospheric refraction along the light path is neglected.

The temperature-pressure (T-P) profile and the molecular volume density profiles used in the model are shown in Figure \ref{TPprofile}. The T-P profile and dewpoint temperature data (which is used to calculate the water vapour pressure) are taken from the MODIS Atmosphere\footnote{Http://modis-atmos.gsfc.nasa.gov} while the ozone and $\mathrm{NO_2}$ data are taken from  the IUP/IFE-UB Sciamachy data\footnote{Http://www.iup.uni-bremen.de/sciamachy/}.
All these atmospheric profiles are carefully chosen from satellites database, so that they represent the properties of the atmosphere  at the positions close to the coordinates shown in Figure \ref{coordinate} and at the time when the lunar eclipse happened. The $\mathrm{O_2}$  volume density is  calculated  with a mixing ratio of $21\%$.

With this geometry and the atmospheric profiles, we can model the transmission spectrum of the light path with different minimum altitude $z_\mathrm{0}$.
By applying the model to the observed spectrum, we get the $z_\mathrm{0}$ from the best-fit model.
The best-fit $z_\mathrm{0}$ is regarded as the ``effective altitude'', which means that the light coming from the path with $z_\mathrm{0}$ determinates the corresponding property of the observed transmission spectrum. In the following, the model spectrum of each atmosphere species is calculated independently.



\subsection{Model calculation of lines}
\subsubsection{$O_2$ column density and oxygen isotopes}
With the model described above, we calculated $\mathrm{O_2}$ transmission spectra for different atmospheric layers with the minimum altitude $z_\mathrm{0}$. Then, we fit the model spectrum to the observed spectrum with the least-squares method: the best-fit  altitude is 12.5 km where the temperature is 218 K and the pressure is 0.18 atm (where atm is the standard atmospheric pressure). 

The overall $\mathrm{O_2}$ column density is determined to be $\mathrm{5.4\times10^{25}\,molecule/cm^2}$ with an uncertainty of $1\%$. 
This uncertainty represents the statistical error from the model fitting and its low value indicates the robustness of the fit. 
However, the result also depends on the assumed theoretical model,  which has its own uncertainty due to the input cross-section data and the geometry approximation. The model uncertainty is not included in our error estimation.

\begin{figure*}
   \centering
   \includegraphics[width=0.9\textwidth]{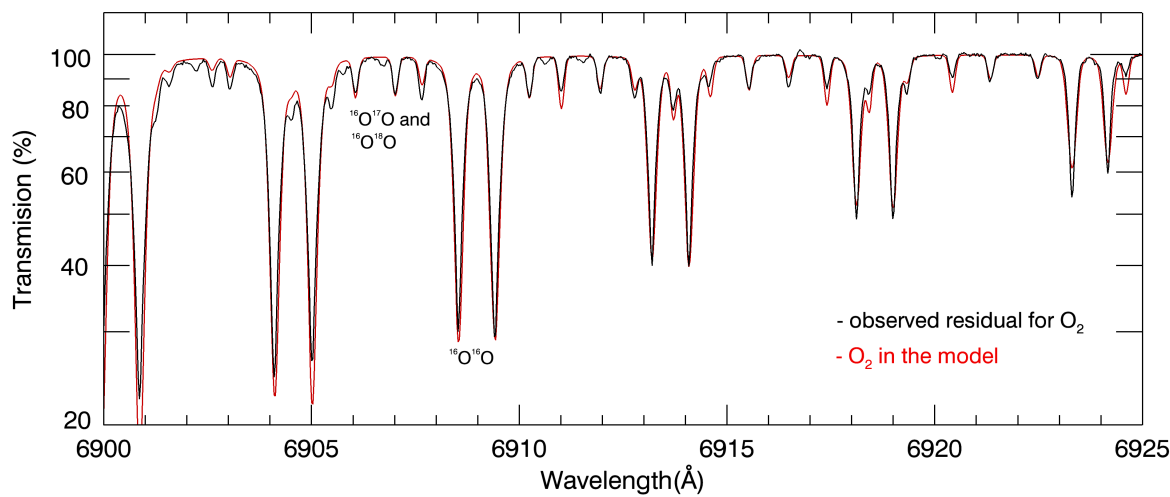}
\caption{The $\mathrm{O_2}$ line model and $\mathrm{O_2}$ observed residual spectrum. The $\mathrm{O_2}$ ``observed residual spectrum'' here means the residual ratio between the observed transmission spectrum and the model spectrum without $\mathrm{O_2}$. This observed residual spectrum can be regarded as a normalised spectrum. The same method is applied to the observed residual spectra of $\mathrm{H_2O}$, $\mathrm{NO_2}$ and $\mathrm{O_2 \cdot O_2}$. The strong absorption lines that occur as doublets are the $\mathrm{^{16}O^{16}O}$ lines while the relatively weak lines are the $\mathrm{^{16}O^{17}O}$ and $\mathrm{^{16}O^{18}O}$ lines.}
         \label{o2lines}
   \end{figure*}

In Figure \ref{o2lines}, the model spectrum of oxygen isotopes  $\mathrm{^{16}O^{16}O}$, $\mathrm{^{16}O^{17}O}$ and $\mathrm{^{16}O^{18}O}$ is plotted, along with the observed spectrum. The $\mathrm{O_2}$ isotope abundances in the normal Earth's atmosphere are employed in our model, i.e. 99.53\% for $\mathrm{^{16}O^{16}O}$, 0.40\% for $\mathrm{^{16}O^{18}O}$ and 0.07\% for $\mathrm{^{16}O^{17}O}$.


%

   \begin{figure*}
   \centering
   \includegraphics[width=0.9\textwidth]{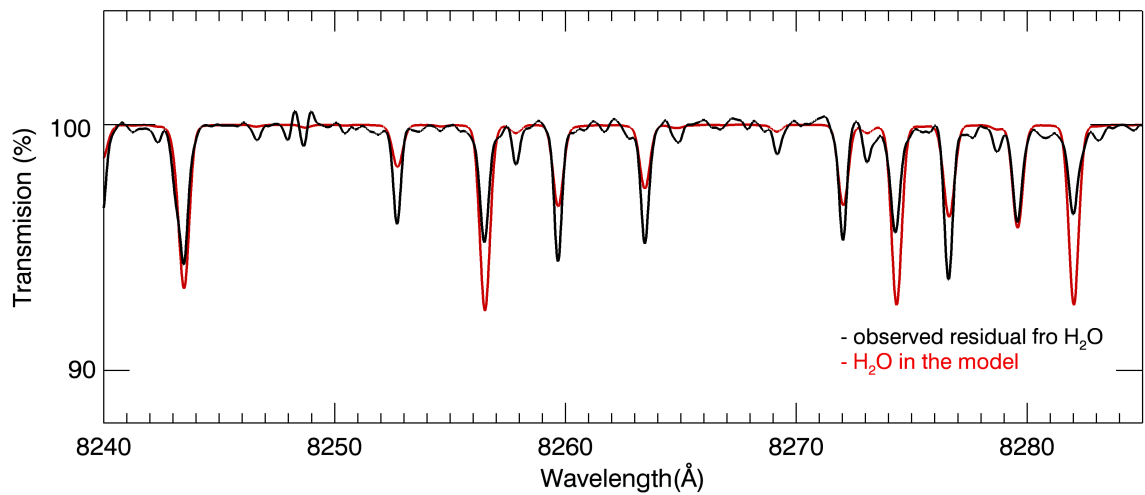}
\caption{The $\mathrm{H_2O}$ model spectrum and the observed residual spectrum. Data are binned every 20 pixels.}
         \label{H2Olines}
   \end{figure*}
%

%
   \begin{figure*}
   \centering
   \includegraphics[width=0.9\textwidth]{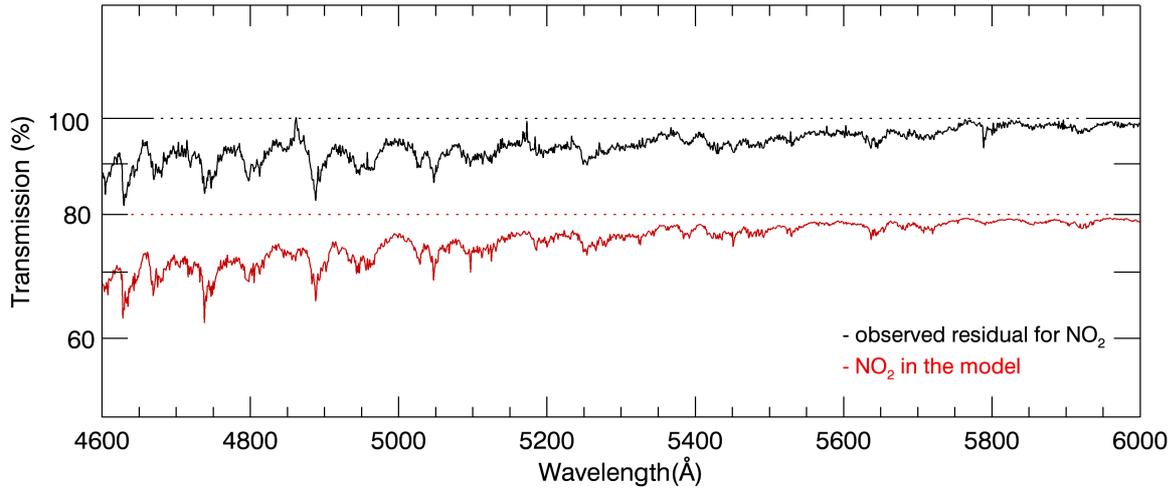}
\caption{The $\mathrm{NO_2}$ model spectrum and the observed residual spectrum. The spectra are binned every 50 pixels. The model spectrum is shifted down 20\% for clarity.}
         \label{NO2}
   \end{figure*}
%
   \begin{figure*}
   \centering
   \includegraphics[width=0.9\textwidth]{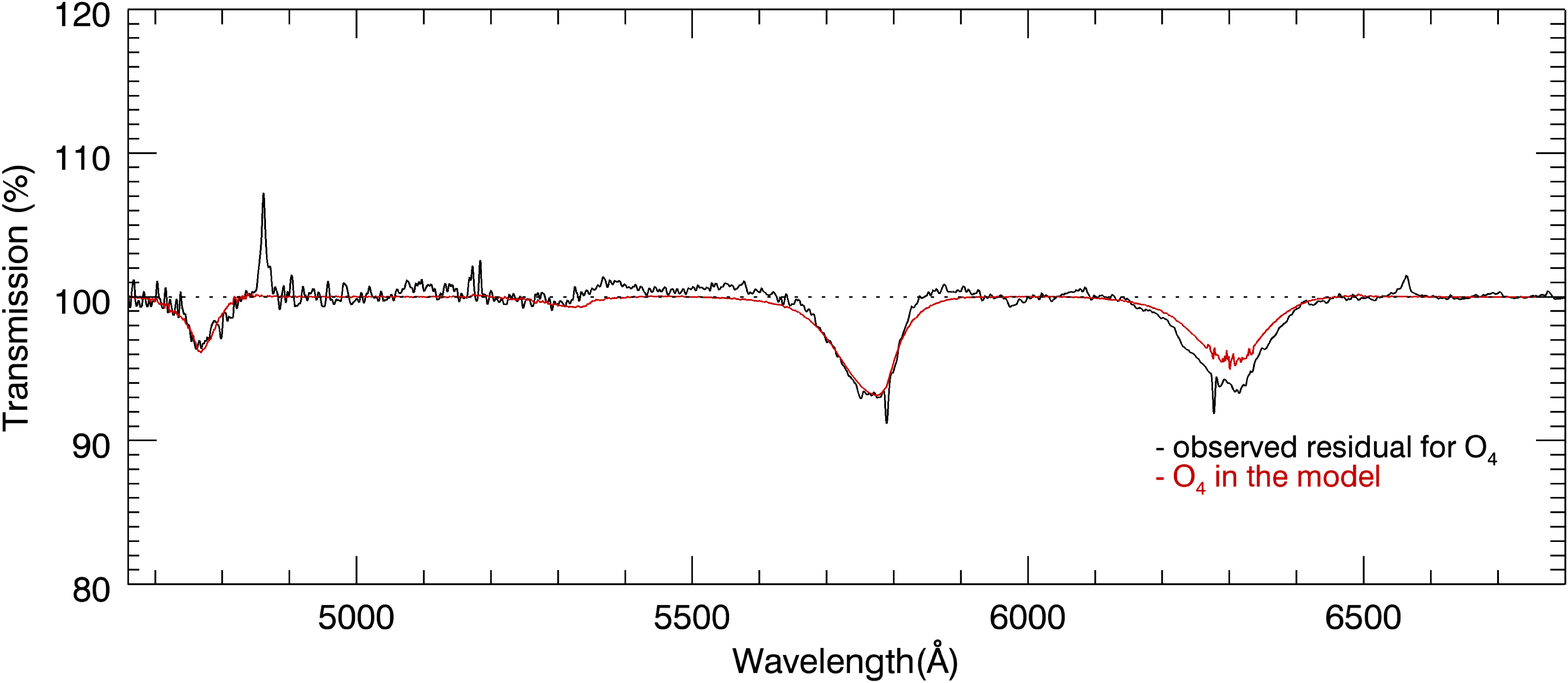}
\caption{The $\mathrm{O_2 \cdot O_2}$ model spectrum and the observed residual spectrum. Data are binned every 300 pixels.}
         \label{O4}
   \end{figure*}
%

\subsubsection{$H_2O$ quantitative calculation}
For the water vapour absorption, we use the cross section at the same temperature and pressure as calculated for the $\mathrm{O_2}$. The observed spectrum is then fitted with the modelled $\mathrm{H_2O}$ transmission spectrum with different column densities.
Figure \ref{H2Olines} shows the best-fit spectrum with the $\mathrm{H_2O}$ column density of $\mathrm{2.0\times10^{21}\,molecule/cm^2}$ with a statistical uncertainty of $2\%$.


The model fit of the water vapour lines is not as good as that of the $\mathrm{O_2}$ lines, mainly because the water absorption in the transmission spectrum is very weak while the water absorption in the vertical absorption above the observatory is relatively strong and even saturated.
Since the transmission spectrum is derived from the ratio of U1 and B2 which were obtained at slightly different vertical airmasses, the water absorption in the ratio spectrum retains a small contaminating component from the vertical absorption.
This small vertical $\mathrm{H_2O}$ component makes the above calculated column density about one third smaller than the actual horizontal $\mathrm{H_2O}$ column density.


\subsection{Model calculation of the continuum}

\subsubsection{Scattering calculation}
Scattering is calculated by Equations (2) and (3). In the model spectrum in Figure \ref{final-model}, the air molecular column density for Rayleigh scattering is $\mathrm{2.42\times10^{26}\,molecule/cm^2}$ with a statistical uncertainty of $0.3\%$, which coincides with the minimum altitude of 13.0 km and a pressure of 0.16 atm. The aerosol factor $A_\mathrm{aerosol}$ is 1.36, suggesting that the aerosol amount is 1.36 times as large as that in the `clear' air condition.

\subsubsection{Ozone calculation}
As indicated by \cite{Orphal2003}, the ozone cross section has only a small temperature dependence within the wavelength range of 4300 - 10000$\,\AA$. Therefore we take the cross section at 210 K for the model calculation.
The column density used in the best-fit model in Figure \ref{final-model} is $\mathrm{3.36\times10^{20}\,molecule/cm^2}$ with an uncertainty smaller than $0.1\%$

From Figure \ref{species} it can be seen that around 6000 $\mathrm{\AA}$ the ozone absorption is stronger than the scattering extinction and thus dominates the transmission spectrum. Indeed, ozone has the highest partial pressure in the atmosphere between 20 km and 25 km, which can result in the eclipsed Moon appearing blue at the edge of the Earth's umbra because of the strong absorption of orange light. A detailed explanation and a photograph of such a blue Moon can be found in \cite{Gedzelman2008}.

\subsubsection{$NO_2$ calculation}
\cite{Voigt2002} found that temperature and pressure affect the cross section of $\mathrm{NO_2}$, however , this is not prominent enough to affect our model calculation.
Therefore, we use the cross section at 223 K and 100 mbar.
In Figure \ref{NO2}, the best-fit model with the $\mathrm{NO_2}$ column density of $\mathrm{3.77\times10^{17}\,molecule/cm^2}$ is shown along with the observed spectrum. The statistic uncertainty of the $\mathrm{NO_2}$ column density is calculated to be $0.1\%$.



\subsubsection{Oxygen CIA calculation}
Using the geometric model described above, the optical depth of the $\mathrm{O_2 \cdot O_2}$ collision-induced absorption along the light path can be expressed as:
\begin{equation}
\tau=\int n_z(\mathrm{O_2})\cdot n_z(\mathrm{O_2}) \cdot \sigma(\mathrm{O_4})\, \mathrm{d} z  ,
   \end{equation}
where $n_z(\mathrm{O_2})$ is the $\mathrm{O_2}$ volume density at altitude $z$ and $\sigma(\mathrm{O_4})$ is the $\mathrm{O_2 \cdot O_2}$ absorption cross section.
This optical depth is then used to calculate the transmission spectrum of $\mathrm{O_2 \cdot O_2}$ for different minimum altitudes $z_\mathrm{0}$. For the best fit spectrum shown in Figure \ref{O4}, the altitude used is 12.3 km, corresponding to a pressure of 0.18 atm and an $\mathrm{O_2 \cdot O_2}$ integrated column density of $\mathrm{5.6 \times10^{43}\,molecule^2/cm^5}$ with a statistical uncertainty of $1\%$. However, with this minimum altitude, the 630 nm band does not fit as well as the 477 nm and 577 nm bands. This may be due to several possible reasons: the combination of $\mathrm{O_2 \cdot O_2}$ and $\mathrm{O_2}$ absorption at 630 nm which makes the measurement of $\mathrm{O_2 \cdot O_2}$ not accurate, the use of improper cross-section for $\mathrm{O_2 \cdot O_2}$ which is measured at 1 atm condition while the absorption of our spectrum happens at 0.18 atm, and the effect of the scattered sunlight.


\subsection{Effective altitudes}

From the above model calculation, we have retrieved the effective altitudes by fitting the $\mathrm{O_2}$ absorption lines, the $\mathrm{O_2 \cdot O_2}$ absorption bands and the molecular scattering independently, i.e. 12.5 km from $\mathrm{O_2}$ model-fit, 12.3 km from $\mathrm{O_2 \cdot O_2}$ and 13.0 km from scattering.

Here we use an effective altitude of 12.5 km --- corresponding to an effective pressure of 0.18 atm --- to calculate the column densities of $\mathrm{H_2O}$, ozone and $\mathrm{NO_2}$ at the same effective altitude with the satellite data shown in Figure \ref{TPprofile}.
The $\mathrm{H_2O}$ column density at this effective altitude is $\mathrm{2.4\times10^{21}\,molecule/cm^2}$ which is consistent with the observed column density of $\mathrm{2.0\times10^{21}\,molecule/cm^2}$.
The corresponding ozone column density is $\mathrm{2.5\times10^{20}\,molecule/cm^2}$ which is a bit smaller than the observed column density of $\mathrm{3.36\times10^{20}\,molecule/cm^2}$.
Given the satellite data error  of several tens of percent and the level of  approximation in our model, the satellite and our observed data can be regarded as consistent with each other with respect to $\mathrm{H_2O}$ and $\mathrm{O_3}$.
However, the $\mathrm{NO_2}$ column density at this effective altitude is $\mathrm{1.4\times10^{17}\,molecule/cm^2}$ -- only $\sim1/3$ of the value of $\mathrm{3.77\times10^{17}\,molecule/cm^2}$, which is  deduced from the observed transmission spectrum. 
The obvious discrepancy is very likely due to the diurnal variation of $\mathrm{NO_2}$.
Its concentration is higher during nighttime as compared to daytime, and peaks at sunset  \citep{Amekudzi2008}.
As the $\mathrm{NO_2}$ SCIAMCHY data used here are taken during local daytime while the lunar eclipse data are necessarily taken at the Earth's terminator, it is reasonable that our $\mathrm{NO_2}$ column density is larger than that of SCIAMCHY data.

\section{Discussion}
\subsection{Water vapour detectability for Earth and exo-Earths}
\begin{figure*}
   \centering
   \includegraphics[width=0.9\textwidth]{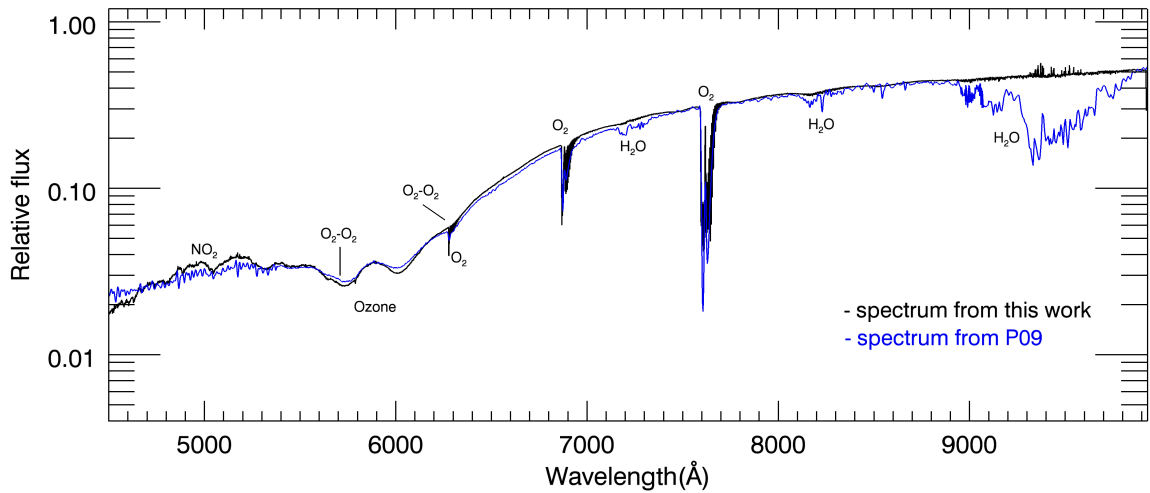}
\caption{
A comparison between our retrieved transmission spectrum, and that from P09. Here, our data have been binned every 50 pixels to adjust to the spectral resolution of P09.
}
         \label{palle}
   \end{figure*}

In our transmission spectrum, the water vapour absorption lines are weak compared with the water absorption in Palle's lunar eclipse observation (see Figure \ref{palle}). The difference is due to the different geometry and different location of the observed atmosphere at the corresponding eclipses. For our spectrum, the location was above the summer Antarctic ocean where the temperature is low and the atmosphere is dry. For Palle's spectrum, the location was above the summer Arctic where the water vapour concentration is higher than that of the summer Antarctica. 
Also, Palle's observing position on the lunar surface is closer to the umbra's center than ours, which means that they probed a lower part of the Earth's atmosphere where the water vapour is more abundant. 
This significant water vapour difference between the two observations indicates a large variability in water vapour detectability which is strongly tied to geometry.



Normally the concentration of water vapour in the Earth's atmosphere drops quickly as the altitude increases. The normal mixing ratio of water vapour at sea level can be $10^4$ ppm but drops below 10 ppm at an altitude of 15 km. For an Earth-like exoplanet it is likely that water vapour will be similarly confined to low altitudes.
According to the atmospheric refraction study by \cite{Garcia2012}, during the mid-transit of an exoplanet system like the Earth-Sun, the atmosphere below 12 -- 14 km can not be observed due to the refraction of stellar light.
Also, clouds in the Earth's atmosphere can act as an optically thick layer and block the atmospheric features below the cloud layer.
Thus the detection of water vapour for an Earth-like exoplanet via transmission spectroscopy will not necessarily be an easy task even if the planetary surface is mainly covered with liquid water or ice. The detection will be even harder if the average atmospheric temperature is low, like that of Antarctica, which will make the atmosphere relatively dry. Detailed transmission models of an exo-Earth's atmosphere including clouds, refraction and different temperature profiles are needed to better constrain the detectability of water vapour.



%

\subsection{$NO_2$ sources}

In our transmission spectrum, $\mathrm{NO_2}$ can be seen obviously, and its absorption is the main spectral feature at the blue end. In the overall Earth's atmosphere,  $\mathrm{NO_2}$ is mainly produced by human activities such as vehicle exhaust. However, the $\mathrm{NO_2}$ observed in our spectrum -- which is located at the low stratosphere above the Antarctica -- has diverse sources.

For the $\mathrm{NO_2}$ in the summer stratosphere, the main source is nitrous oxide ($\mathrm{N_2O}$) oxidation by the excited oxygen atoms \citep{Crutzen1979}. What is interesting is that $\mathrm{N_2O}$ is produced by microbial oxidation-reduction-reactions and is regarded as a biosignature \citep{Seager2010}. Thus here we point out that the detection of $\mathrm{NO_2}$ in the stratosphere can be regarded as a possible hint for the existence of the biosignature gas -- $\mathrm{N_2O}$.

$\mathrm{NO_2}$ can also be transported from troposphere to stratosphere. For summer troposphere above the Antarctica, $\mathrm{NO_2}$ comes from human activities as well as from the snowpack melting \citep{Jones2001}. 

Mt. Erebus, a volcano with a lava lake which has been active for many decades, is located close to our transmission spectrum trajectory (the position is shown in Figure \ref{coordinate}). As indicated by \cite{Oppenheimer2005}, Erebus volcano may be a significant source of the  $\mathrm{NO_2}$ in Antarctic atmosphere. This implies $\mathrm{NO_2}$ may be detectable for a geological active exoplanet which may be pumping large amount of $\mathrm{NO_2}$ into the atmosphere. Detailed geology and chemistry study will be required for this assumption.



\subsection{Skylight contribution}

\begin{figure*}
   \centering
   \includegraphics[width=0.9\textwidth]{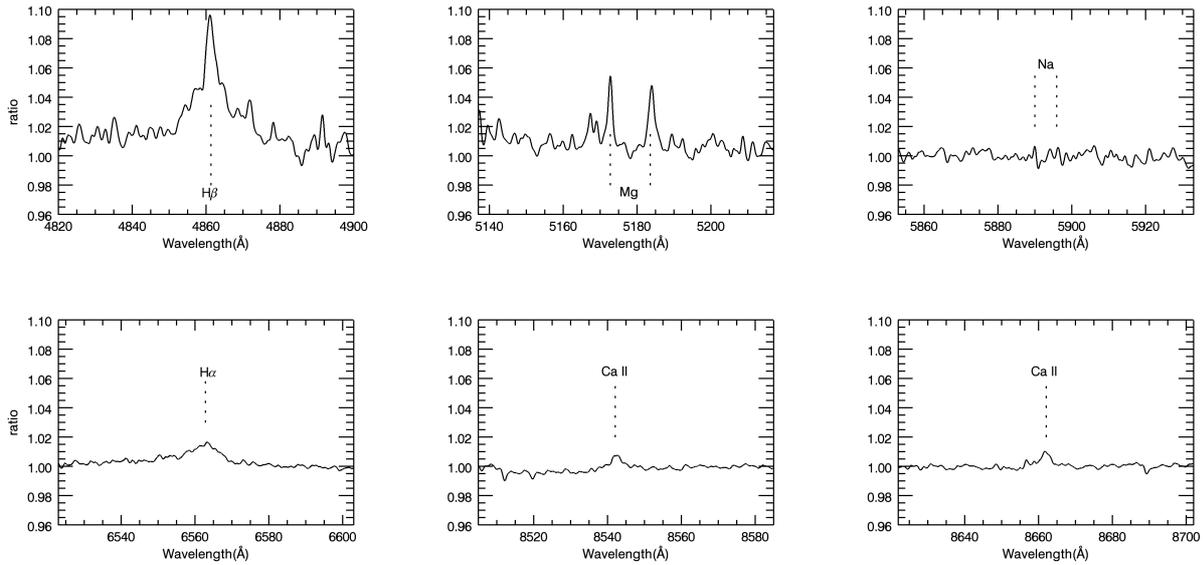}
\caption{`Emission features' of the solar Fraunhofer lines in the ratio of U1 to B2. Here we display the residual of the observed spectrum and our modelled transmission spectrum to remove all the transmission spectrum features. A radial velocity correction of 0.4 km/s is applied to eliminate the RV structure of the solar lines as discussed in Section 3.3.}
         \label{Raman}
   \end{figure*}

In the model described above, we have only considered the transmitted sunlight. However, the sunlight scattered by the atmosphere at the Earth's terminator (i.e. terminator skylight) also contributes to the brightness of the eclipsed Moon \citep{Garcia2011a}. In this section we show the effect of the Raman scattering \citep[the Ring Effect, ][]{Grainger1962} which modifies the skylight spectrum and allows us to demonstrate the existence of scattered sunlight in the umbra spectrum (see Figure \ref{Raman}). These Raman scattering features have never been shown before in the previous umbral lunar eclipse observations.

In the residual of the observed spectrum and our model spectrum, there are apparent `emission' features in the positions of the strong Fraunhofer lines. These indicate that the solar Fraunhofer lines in U1 are shallower than in B2, resulting in an apparent emission in the ratio spectrum.
This filling-in effect at the Fraunhofer lines is called Ring Effect and was initially discovered when comparing the diffuse skylight spectrum with a direct lunar spectrum.
This effect in the skylight spectrum is believed to be  caused by Raman scattering that transfers continuum photons into the solar absorption lines, making them shallower than in the original solar spectrum \citep{Langford2007}.
Since our directly transmitted sunlight will not suffer the filling-in effect, we propose that these apparent `emission' features are caused by  a significant fraction of scattered sunlight appearing in the umbra spectrum.

%
%

\begin{figure*}
   \centering
   \includegraphics[width=0.9\textwidth]{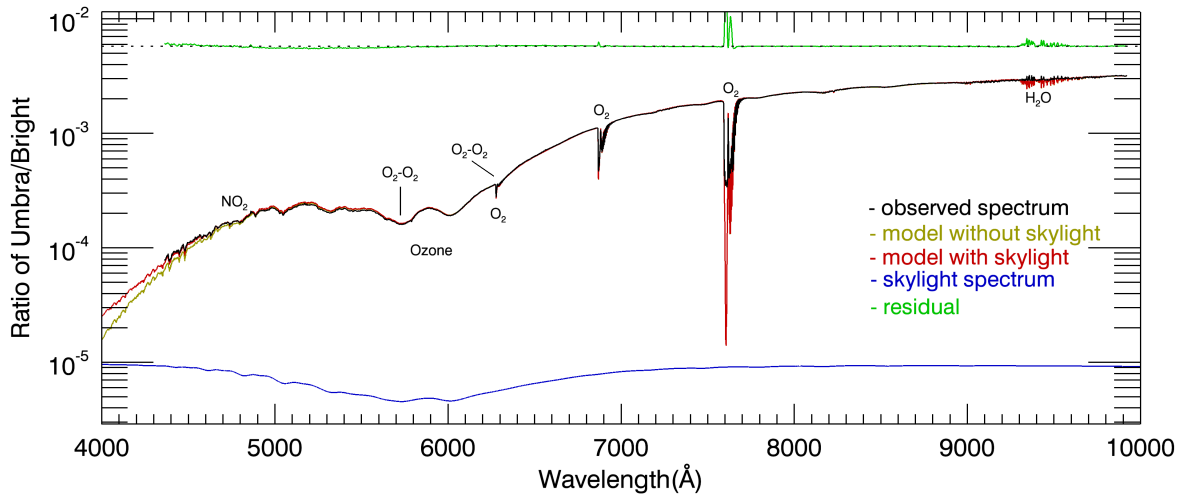}
\caption{The skylight model spectrum. The blue line is the skylight model and the yellow line is the transmission model while the red line is the transmission model added by the skylight model. The black line is our observed spectrum and the green line shows the residual of observed spectrum and model spectrum with the skylight. }
         \label{skylight}
   \end{figure*}

Although these `emission' features are relatively weak, they can still be clearly identified in the residual spectrum. In Figure \ref{Raman}, almost all the strong solar Fraunhofer lines display the emission shape, such as $\mathrm{H_\beta}$, Mg lines, $\mathrm{H_\alpha}$, Ca\,II lines. The `emission' features appear stronger at the short wavelength range and $\mathrm{H_\beta}$ displays the strongest emission feature. This is because the transmitted sunlight is weak at short wavelengths due to strong Rayleigh and aerosol scattering so that the scattered light becomes prominent there.
However, as shown in the top left figure of Figure \ref{Raman}, we can barely find any emission features for the Na D lines as other strong Fraunhofer lines, which is possibly a result from the combination of the strong ozone absorption around Na D lines in the skylight (see the modelled skylight spectrum in Figure \ref{skylight}) and Na resonance absorption in the Earth's atmosphere.

Following the method provided by \cite{Link1972}, we built a simple skylight model for the atmospheric ring at the Earth's terminator based on \textit{the U.S. Standard Atmosphere 1976}\footnote{Http://modelweb.gsfc.nasa.gov/atmos/us$\_$standard.html}.
This skylight model is used together with the transmission model to fit the observed spectrum, and the result is shown in Figure \ref{skylight}. The y axis for the observed spectrum here is the ratio of U1 and B2 spectra divided by the integration time ratio of 300. Thus this y axis can be regarded as the flux which is normalised to the direct solar spectrum as received at the lunar surface.
For the best-fit model shown in Figure \ref{skylight}, we used the same transmission model as in Figure \ref{final-model}, adding a skylight model with a strength around $10^{-5}$ at $4500\,\AA$.
The residual in the figure reveals the model fits quite well at the blue end compared with the model without skylight. We emphasize here that the quantitative amount of the skylight calculated above is not robust because of model-fit degeneracies. Detailed model of the Ring Effect together with the skylight spectral model is necessary for accurate determination of skylight contribution.


%

%

\subsection{Comparing with other lunar eclipse observations}
It needs to be emphasized that our transmission spectrum is of high resolution and high SNR, which is the major improvement compared to previous studies. 
Such observations provide more detailed results of the Earth's transmission spectrum, and lead to clear detections of $\mathrm{O_2}$ isotopes and the Raman scattering, which have never been done in previous lunar eclipse observations.
Below we show several specific differences comparing with previous observations.

From our high resolution and high SNR spectrum, no absorption features from atoms or ions are detected. Here we point out that the detection of Ca T (Ca\,II near infrared triplet lines) in the Earth's atmosphere is highly unlikely because they are not absorption lines from the ground state of Ca\,II. 
Thus the unexpected presence of Ca\,II in P09's spectrum was probably not due to the absorption of Ca\,II in the Earth's atmosphere.


Given that we observed the same lunar eclipse (10 December 2011) as U13 and the 
observing locations on the lunar surface are similar (i.e. around the Tycho crater), it is worthy to compare the two results quantitatively.
In general our results are consistent with U13's for the strong absorption species such as $\mathrm{O_2}$ and $\mathrm{O_3}$. However, for the weak absorption species, i.e. $\mathrm{NO_2}$, $\mathrm{H_2O}$ and $\mathrm{O_2 \cdot O_2}$, there are differences between the two observations. The $\mathrm{NO_2}$ column density calculated by us is twice of that calculated by U13.
The water vapour column density in U13 is calculated in a different way and is about 3 times larger than our calculated $\mathrm{H_2O}$ column density. U13 shows that the retrieved effective altitude at the 477 nm $\mathrm{O_2 \cdot O_2}$ CIA band is much lower than that at the 577 nm band and attributes the result to the lower path of scattered sunlight at 477 nm band, however, the effective altitude of these two CIA bands are almost the same in our results.



\cite{Garcia2011b} analysed $\mathrm{O_2 \cdot O_2}$ 577 nm and 630 nm bands of P09 spectrum and found that the integrated column density at 630 nm is about twice the column density at 577 nm. 
They further invoked scattered sunlight in the umbra spectrum as the main explanation. As the lunar eclipse of P09 is volcanically perturbed, their umbra spectrum may have more scattered sunlight due to aerosols \citep{Garcia2011a}.
Our result shows that the column density of 630 nm is $\sim$ 1.5 times of the 577 nm, this may be partly due to the scattered sunlight. However, we do not attribute this mainly to the scattered sunlight because our lunar eclipse is not volcanically perturbed, which means the amount of the scattered sunlight should be significantly smaller than that of P09. Further observation and model calculation are needed to better understand these two $\mathrm{O_2 \cdot O_2}$ CIA bands in the transmission spectrum.


\section{Conclusions}

%
%

We have shown and discussed the high resolution and high SNR transmission spectrum of the Earth's atmosphere observed during a total lunar eclipse. 
The effective absorption altitude for the observed spectrum is found to be around 12.5 km, which is derived by combining the scattering, $\mathrm{O_2}$ and $\mathrm{O_2 \cdot O_2}$ model calculations. The column densities of $\mathrm{O_3}$, $\mathrm{H_2O}$ and $\mathrm{NO_2}$ are compared with satellite data at this altitude. 
In the transmission spectrum, the most prominent absorption feature is the Chappuis band of ozone which will be an important feature for future exo-Earth characterizations. The $\mathrm{O_2}$ absorption lines are resolved, and the different oxygen isotopes are clearly detected for the first time in a lunar eclipse observation.
The water vapour absorption is found to be weak in our spectrum comparing with previous lunar eclipse observations because the atmosphere we probed is relatively dry. This indicates the large variability in water vapour detectability in an Earth-like exoplanet’s atmosphere via transmission spectroscopy.
Further model studies including different water vapour mixing ratio profiles, the atmospheric refraction and the water cloud are needed to constrain the water vapour detectability.
The observed $\mathrm{NO_2}$ column density is significantly higher than the value obtained from the satellite data but this is attributed to the the diurnal variation of $\mathrm{NO_2}$ concentration caused by $\mathrm{NO_2}$ photochemistry. This demonstrated the importance of photochemistry process in transmission spectroscopy.
We also detected the effects of Raman scattering in the strong Fraunhofer lines (the Ring Effect) --- with the exception of the Na D-lines --- which can be regarded as an indicator for forward-scattered sunlight reflected from the Moon.


For the three unique biosignature gases of Earth ($\mathrm{O_3}$, $\mathrm{O_2}$, $\mathrm{N_2O}$) \citep{Lovelock1965,Seager2012}, $\mathrm{O_3}$ and $\mathrm{O_2}$ show strong features in our spectrum while the presence of $\mathrm{NO_2}$ in the stratosphere indicate the existence of $\mathrm{N_2O}$ indirectly.

Detailed studies of the transmission spectra obtained during lunar eclipses, provide not only a chance to study the Earth atmosphere, but also the excellent guidance for the studies of the atmospheres of terrestrial exoplanets. 
Our transmission spectrum can be used as a comparison for more complex terrestrial atmospheric models, which will yield better estimates of the detectable features for exo-Earth characterization.
Further observations of lunar eclipses could provide a set of transmission spectra at various altitudes and at various weather/climate conditions.


\begin{acknowledgements}
This research is based on data collected at Xinglong Station, which is operated by National Astronomical Observatories, CAS, and is supported by the National Natural Science Foundation of China under grants 11233004, 11173031, 11203035 and 11390371.
The authors warmly thank the reviewers for helpful reviews, the European Space Agency and IUP/IFE-UB for the SCIAMACHY data and Clive Oppenheimer and Petra D'ODorico for useful discussions.
F. Yan acknowledges the support from ESO-NAOC studentship.
\end{acknowledgements}

\bibliographystyle{aa} 

\bibliography{Eclipsepaper}


\end{document}